\documentclass[runningheads,a4paper]{iaga}

\usepackage{amssymb}
\usepackage{graphicx}

\usepackage{url}
\newcommand{\keywords}[1]{\par\addvspace\baselineskip
\noindent\keywordname\enspace\ignorespaces#1}

\begin{document}

\mainmatter  

\title{Daytime seeing and solar limb positions}

\titlerunning{Daytime Seeing Sigismondi}

\author{Costantino Sigismondi}

\authorrunning{C. Sigismondi}

\institute{ICRA-Sapienza, P.le Aldo Moro 5 00185 Rome Italy; \\
University of Nice-Sophia Antipolis, Dept. Fizeau UMR 6525 Nice France; \\
IRSOL, via Patocchi, 6610 Locarno Monti, Switzerland\\
\email{sigismondi@icra.it}
\\ home page:
\texttt{http://www.icra.it/solar}}


\maketitle

\begin{abstract}
A method to measure the seeing from video made during drift-scan solar transits is proposed.
The limb of the Sun is projected over a regular grid evenly spaced. The temporal dispersion of 
the time intervals among the contacts between solar limb and grid's rows is proportional to the atmospheric seeing.
Seeing effects on the position of the inflexion point of the limb's luminosity profile are calculated numerically with Fast Fourier Transform.
Observational examples from Locarno and Paris Observatories are presented to show the asymmetric contributions of the seeing at the beginning and the end of each drift-scan transit. 
\end{abstract}

\section{Introduction}

The atmospheric turbulence produces the well known effects of blurring and image motion, animated by different timescales. With a single parameter $r_0$ it is possible to describe this phenomenon\cite{fried1,fried2}, while the atmospheric turbulence can be investigated more deeply considering the scale heights and dimensions of turbulent cells\cite{schoeck}, to select the best observational sites\cite{arena}.

\noindent A formula relating the wavelength of light $\lambda$, the seeing $\rho$ and $r_0$ is: 

$\rho$["]= $\lambda$ [nm]/$5.5\cdot r_0$ [mm]

\noindent The seeing is the full width half maximum FWHM of the Point Spread Function PSF observed through turbulent atmosphere (being the telescope's objective larger than $r_0$), or, conversely, it can be considered as the diameter of a telescope with diffraction limit equal to $\rho$. 

A simple rule for $\lambda$ of visible light is $r_0$=100 mm /$\rho$ ["].

The need of measuring daytime seeing in real time, arises from the fact that atmospheric turbulence modifies the perceived position of the solar limb in long-exposure images. The concept of long-exposure is relative to seeing's timescales, and not to the quantity of photons detected. In practical cases we can consider long-exposure a duration $\tau \ge 0.02$ s, which is always larger than usual exposure times from modern commercial videocameras (CCD and CMOS) set with authomatic gain ($\tau \sim 0.01$ s). The abundance of photons makes possible to perceive the degradation of limb's profile in such a short time.

\section{The drift-scan method}
The tracking system of the telescope is stopped and the telescope is in the drift-scan mode: the Sun moves on the fixed field of view of the telescope and sweeps on it. 
The drift-scan method for measuring the solar diameter\cite{witt1,witt2} consists to identify 
the contact times of the preceding and following limb of the Sun on a given hourly circle. The duration of this
transit is proportional to the solar diameter $\Delta T\propto\emptyset_{\odot}\cdot cos(\delta_{\odot})$ with further corrections for $\dot\delta_{\odot}$\cite{WittA,WittB}.
Typical transits durations exceed 120 s, and during this period the atmospheric turbulence along the line of sight can change. The seeing at the first contact can be different from the value at the last contact. And this seeing effect can further shift the position of the inflexion points of the luminosity profile (they are conventionally considered as the position of the solar limb\cite{hill}) asymmetrically for the two limbs.
A method to evaluate in real time this parameter is here presented, as well as a numerical estimate of the seeing effect on the measurement of solar diameter. 

\section{Daytime seeing}
The most simple way to measure the seeing is by projection of the solar image on a regular grid during a drift-scan observation.
A videocamera records the transit of the solar limbs above the grid, and the time intervals 
required to cover the evenly spaced intervals of the grid are measured by a frame by frame inspection.
I used a SANYO CG9 videocamera recording at 60 frames per second on mp4 format and I analyzed these video with QUICKTIME 7.0 freeware.  
Without any seeing effect the uniform speed of the solar limb would produce a series of contact times separated by equal time intervals, while the turbulence produces a scatter in these data. The standard deviation of these time intervals $\sigma$ [s] is related to the seeing $\rho$ ["] by the approximate formula

$\rho$=$\sigma \cdot$ 15$ \cdot$ cos($\delta_{\odot}$)

\noindent where $\delta_{\odot}$ is the declination of the Sun at the moment of the observation.

This formula is accurate at $1\%$ level, enough for these studies. 
As examples of this method I can report the observations made by me and Michele Bianda the 45 cm Gregorian Telescope of IRSOL, located in Locarno, Switzerland, and these made by Cyril Bazin at the 33 cm Carte du Ciel Astrograph, located in central Paris at the astronomical observatory founded by Giandomenico Cassini.

Paris       17 April 2010 h. 10: $\rho_1$=0.79" $\rho_2$=1.28" 

Paris        24 April 2010 h. 9: $\rho_1$=1.23" $\rho_2$=0.86" 

Locarno     9 August 2008 h. 17: $\rho_1$=0.76" $\rho_2$=0.60" 

Locarno 9 August 2008 h. 17 (2): $\rho_1$=0.87" $\rho_2$=0.95" 

First and second contact seeing evaluations are labelled with 1 and 2. 

\noindent The turbulence during the transits observed in Paris in the morning, is greater than in Locarno during afternoon, thanks to special conditions of the atmosphere near the lake Maggiore.
The asimmetry for the two contacts is always present.

\section{Numerical calculation of limb's shift due to the seeing}

Hill, Stebbins and Oleson\cite{hill} consider the inflexion point of the luminous profile of the Sun as an operatively stable definition of solar limb. The stability is relative to different seeing conditions. Using a standardized Limb Darkening Function\cite{roger} LDF and a gaussian PSF as representative of the seeing effect I have calculated the effect of the convolution
of LDF with PSF respectively corresponding to $\rho$=1", 2" and 3".
The results in solar diameter's perceived shrinking have been respectively $\Delta\emptyset_{\odot}$=-0.07", 0.17" and 0.28". 
Therefore the limb's inwards (toward the center of the disk) shift is respectively 35, 85 and 140 milliarcsec, for $\rho$=1", 2" and 3". 
For the convolution theorem I have used the Fast Fourier Transform FFT algorithm with 1024 data points.
A complete study should take into account the use of opportune filters with given wavelength and bandwidth, since the LDF and the solar diameter are different for different wavelengths\cite{kout,neckel}, within one arcsecond of spread.

\section{Conclusions}
This method of real time measurements of daytime seeing using the progreding image of the Sun above a evenly spaced grid is very simple and easy to be made. It is also suitable for didactic applications, as well as quick site and seeing test for solar observations.

\noindent Test with the pinhole of $\emptyset=1.6$ cm of Santa Maria degli Angeli\cite{SMA} showed a seeing $\rho$=5.6". This large value can be due to the hot air rapidly moving near the pinhole, but this is also the diffraction limit of this small pinhole. Telescopes with $\emptyset\ge30$ cm are never limited by diffraction to measure the seeing with this method, but for all practical uses $\emptyset\ge20$ cm is enough.

\noindent During drift-scan experiments for the measurement of solar diameter it is possible to know the instantaneous values of the seeing $\rho_1$ and $\rho_2$ at the first and second contacts.

\noindent A numerical evaluation of inwards limb's shift as function of the seeing has been presented and it can be used to correct the solar diameter value for seeing $\rho$ and for differential seeing effects $\rho_1 \not=\rho_2$, to exploit the possibility to know the seeing at the moment of the individual contacts of the solar limb with a given hourly circle.

\begin{figure}
\centering
\includegraphics[height=7.5cm]{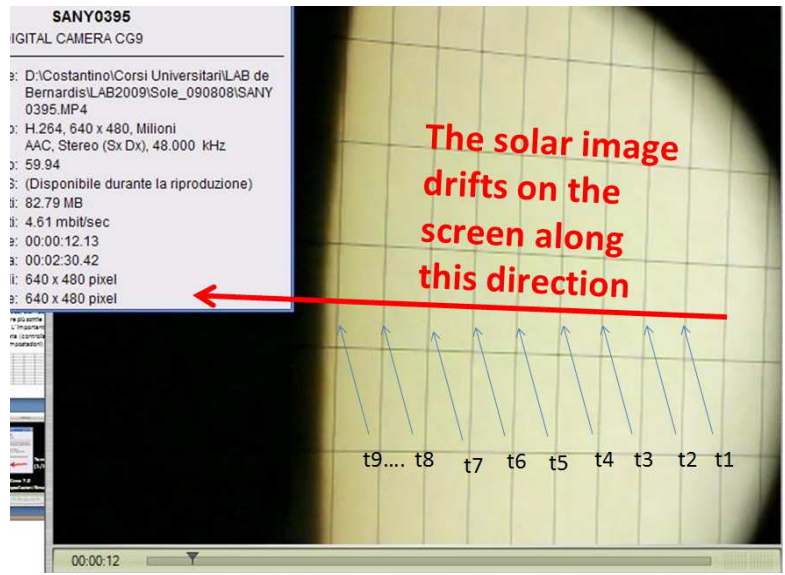}
\caption{Image of a video-frame of the transit of the Sun seen in Locarno on 9 August 2008, the solar image 
is projected on the evenly spaced grid. Time intervals are indicated on this figure. The field of view of that image is about 200 arcsec wide.}
\end{figure}

\subsubsection*{Acknowledgments.} It is a pleasure to record my thanks to Dr. Michele Bianda, director of the IRSOL observatory; to Ing. Cyril Bazin who provided me with the observations at the Carte du Ciel in Paris and to Prof. Serge Koutchmy who set up this experiment at the Observatoire de Paris and who invited me there from April 30 to May 14, 2010.

\end{document}